\documentclass[11pt]{article}
\usepackage{amssymb,amsmath,amsfonts}
\usepackage{graphicx}
\usepackage{graphics}
\usepackage{eepic,epsfig}

\begin{document}

\title{Optical transition radiation in presence of acoustic waves \\
for an oblique incidence}
\author{A. R. Mkrtchyan, V. V. Parazian\thanks{%
E mail: vparazian@gmail.com}, A. A. Saharian\thanks{%
E mail: saharian@ysu.am} \vspace{0.5cm} \\
\textit{Institute of Applied Problems in Physics,}\\
\textit{25 Nersessian Street, 0014 Yerevan, Armenia}} \date{ }
\maketitle

\begin{abstract}
Forward transition radiation is considered in an ultrasonic superlattice
excited in a finite thickness plate under oblique incidence of relativistic
electrons. We investigate the influence of acoustic waves on both the
intensity and polarization of the radiation. In the quasi-classical
approximation, formulas are derived for the vector potential of the
electromagnetic field and for the spectral-angular distribution of the
radiation intensity. It is shown that the acoustic waves generate new
resonance peaks in the spectral and angular distributions. The heights and
the location of the peaks can be controlled by choosing the parameters of
the acoustic wave. The numerical examples are given for a plate of fused
quartz.
\end{abstract}

\bigskip

\textit{Keywords:} Interaction of particles with matter; physical effects of
ultrasonics.

\bigskip

PACS Nos.: 41.60.Dk, 43.35.Sx, 43.35.-c

\section{Introduction}

Transition radiation is produced when a uniformly moving charged particle
crosses an interface between two media with different dielectric properties.
Such radiation has a number of remarkable properties and at present it has
found many important applications (see, for instance, Refs. \cite{TerMik}-%
\cite{Poty09}). In particular, the transition radiation is widely used for
particle identification, for the measurement of transverse size, divergence
and energy of electron and proton beams. An enhancement for the transition
radiation intensity may be achieved by using the interference between the
radiation emitted by many interfaces in a multilayer structure. From the
point of view of controlling the parameters of various radiation processes
in a medium, it is of interest to investigate the influence of external
fields, such as acoustic waves, temperature gradient etc., on the
corresponding characteristics. The considerations of processes, such as
diffraction radiation \cite{MkrtDR}, parametric X-radiation \cite{Mkrt91},
channeling radiation \cite{Mkrt86}, bremsstrahlung of high-energy electrons
\cite{Mkrt02}, electron-positron pair creation by high-energy photons \cite%
{Saha04Brem}, have shown that the external fields can remarkably change the
spectral and angular characteristics of the radiation intensities. Recently
there has been broad interest in compact crystalline undulators with
periodically deformed crystallographic planes as an efficient source of high
energy photons (for a review see Ref. \cite{Koro04}).

In Refs. \cite{Grig98,Mkrt10} we have considered the X-ray and optical
transition radiation from ultrarelativisitic electrons in an ultrasonic
superlattice excited in melted quartz plate. The radiation from a charged
particle for a semi-infinite laminated medium has been recently discussed in
\cite{Grig09}. In these references the transition radiation is considered
under normal incidence of a charged particle upon the interface of the
plate. In the present paper we generalize the corresponding results for
oblique incidence. The angle between the particle velocity and the normal to
the interface is an additional parameter which can be used for the control
of angular-frequency distribution and the polarization of the radiation.

We have organized the paper as follows. In the next section we evaluate the
vector potential for the electromagnetic field by using the quasi-classical
approximation. The intensity of the radiation in forward direction is
investigated in section \ref{sec:Radiation} for both parallel and
perpendicular polarizations. In section \ref{sec:Numeric} we present
numerical examples for the radiation intensity in the case of a plate made
of fused quartz. The main results of the paper are summarized in section \ref%
{sec:Conc}.

\section{Electromagnetic field}

\label{sec:Fields}

We consider the transition radiation under oblique incidence of a charged
particle on a plate with dielectric permittivity $\varepsilon _{0}$ which is
immersed into a homogeneous medium with permittivity $\varepsilon _{1}$. We
assume that the plate has the thickness $l$ and its boundaries are located
at $z=-l$ and $z=0$. In what follows the $z$ axis is directed along the
normal to the plate. The trajectory of the particle with velocity $\mathbf{v}
$ is in the $(x,z)$ plane and forms with the $z$ axis a given angle $\alpha $%
. So, we can write $\mathbf{v}=\left( v\sin \alpha ,0,v\cos \alpha \right) $%
. We assume that longitudinal ultrasonic vibrations are excited in the plate
along the normal to its surface, that form a superlattice. In the presence
of the ultrasonic waves, the dielectric permittivity can be written in the
form
\begin{equation}
\varepsilon \left( z\right) =\left\{
\begin{array}{cc}
\varepsilon _{0}+\Delta \varepsilon \cos \left( k_{s}z+\omega _{s}t+\phi
\right) , & -l\leqslant z\leqslant 0, \\
\varepsilon _{1}, & z<-l,\;z>0,%
\end{array}%
\right. .  \label{epscond}
\end{equation}%
In (\ref{epscond}), $\omega _{s}$, $k_{s}$ are the cyclic frequency and the
wave number of the ultrasound, and $\phi $ is the initial phase. Under the
condition $\nu _{s}l/v\ll 1$, with $\nu _{s}=\omega _{s}/(2\pi )$, during
the transit time of the particle the dielectric permittivity in the
superlattice is not notably changed. For relativistic particles and for the
plate thickness $l\lesssim 1$ cm this leads to the constraint $\nu _{s}\ll
10^{11}$ Hz.

Here we are interested in the radiation with frequencies $\omega $ in the
spectral range $\omega \gg k_{s}c$. The presence of the small parameter $%
k_{s}c/\omega $ allows us to employ the quasi-classical approximation for
the evaluation of the radiation field in the forward direction. It is
natural that in this case the plate boundaries must be sufficiently smooth.
This condition is assumed to be observed \cite{TerMik} because the
transition radiation on these boundaries is formed in a zone with
macroscopic length. For the current density one has the expression
\begin{equation}
\mathbf{j}=e\mathbf{v}\delta \left( x-x\left( t\right) \right) \delta \left(
y\right) \delta \left( z-z\left( t\right) \right) ,  \label{currentmosh}
\end{equation}%
where $e$ is the charge of the particle and
\begin{equation}
x\left( t\right) =v\left( t-t_{0}\right) \sin \alpha ,\;z\left( t\right)
=-l+v\left( t-t_{0}\right) \cos \alpha .  \label{zt}
\end{equation}%
Here we assume that the particle trajectory is rectilinear (for the
discussion of multiple scattering effects see \cite{TerMik}).

In the Lorentz gauge, the vector potential of the electromagnetic field
corresponding to the source (\ref{currentmosh}) can be taken as $\mathbf{A}%
=\left( A_{x},0,A_{z}\right) $. We define the partial Fourier transform as
\begin{equation}
\mathbf{A}\left( \omega ,\mathbf{q},z\right) =\frac{1}{(2\pi )^{3}}\int
dt\int d\mathbf{r}_{\perp }\,\mathbf{A}\left( t,\mathbf{r}\right) e^{i\left(
\omega t-\mathbf{qr}_{\perp }\right) },  \label{vectorpotdet}
\end{equation}%
with $\mathbf{q}=(k_{x},k_{y})$ and $\mathbf{r}_{\perp }=(x,y)$. In the
quasi-classical approximation, these components are determined by the
following expressions \cite{Khachatryan} (see also \cite{TerMik} for the
discussion of applicability of this approximation)
\begin{eqnarray}
A_{x}\left( \omega ,\mathbf{q},z\right) &=&\frac{-i}{2k_{z}^{1/2}(z)}\int
dz^{\prime }\frac{F_{x}\left( z^{\prime }\right) }{k_{z}^{1/2}(z^{\prime })}%
\exp \left( i\int_{z^{\prime }}^{z}dz^{\prime \prime }k_{z}(z^{\prime \prime
})\right) ,  \notag \\
A_{z}\left( \omega ,\mathbf{q},z\right) &=&-\frac{i\sqrt{\varepsilon \left(
z\right) }}{2k_{z}^{1/2}(z)}\int dz^{\prime }\frac{\sqrt{\varepsilon \left(
z^{\prime }\right) }}{k_{z}^{1/2}(z^{\prime })}\exp \left( i\int_{z^{\prime
}}^{z}dz^{\prime \prime }k_{z}(z^{\prime \prime })\right)  \notag \\
&&\times \left[ \frac{F_{z}\left( z^{\prime }\right) }{\varepsilon \left(
z^{\prime }\right) }+\frac{ik_{x}}{\varepsilon ^{2}\left( z^{\prime }\right)
}\frac{d\varepsilon \left( z^{\prime }\right) }{dz^{\prime }}A_{x}\left(
\omega ,\mathbf{q},z^{\prime }\right) \right] .  \label{Akxdef}
\end{eqnarray}%
In (\ref{Akxdef}), we have defined the functions%
\begin{equation}
k_{z}(z)=\sqrt{\frac{\omega ^{2}}{c^{2}}\varepsilon \left( z\right) -q^{2}},
\label{kzz}
\end{equation}%
and%
\begin{equation}
F_{p}\left( z\right) =\mathbf{-}\frac{1}{2\pi ^{2}c}\int dt\int d\mathbf{r}%
_{\perp }\,\,j_{p}\left( \mathbf{r,}t\right) e^{i\left( \omega t-\mathbf{qr}%
_{\perp }\right) },  \label{Fpz}
\end{equation}%
with $p=x,z$.

For the problem under consideration, by using the expressions for the
components of the current density, the functions $F_{p}\left( z\right) $ are
written in the form
\begin{eqnarray}
F_{z}\left( z\right) &=&\mathbf{-}\frac{ee^{i\omega t_{0}}}{2\pi ^{2}c}\exp %
\left[ i\frac{z+l}{v\cos \alpha }\left( \omega -k_{x}v\sin \alpha \right) %
\right] ,  \notag \\
F_{x}\left( z\right) &=&F_{z}\left( z\right) \tan \alpha  \label{FzFx}
\end{eqnarray}%
Assuming that $\Delta \varepsilon $ is sufficiently small, which is well
satisfied for the perturbations induced by the ultrasound, for the function (%
\ref{kzz}) we can write
\begin{equation}
k_{z}(z)=\left\{
\begin{array}{ll}
k_{z}^{\left( 1\right) }, & z<-l,\;z>0, \\
k_{z}^{\left( 0\right) }+ak_{s}\cos \left( k_{s}z+\varphi _{1}\right) , &
-l\leqslant z\leqslant 0,%
\end{array}%
\right.  \label{k3z}
\end{equation}%
with $\varphi _{1}=\omega _{s}t_{0}+\phi $. Here and in what follows we use
the notations
\begin{eqnarray}
k_{z}^{\left( j\right) } &=&\sqrt{\omega ^{2}\varepsilon _{j}/c^{2}-q^{2}}%
,\;j=0,1,\;  \label{defamek} \\
a &=&\frac{\omega ^{2}\Delta \varepsilon }{2c^{2}k_{s}k_{3}^{\left( 0\right)
}}.  \label{a}
\end{eqnarray}%
Note that $a\sim \lambda _{s}\Delta \varepsilon /\lambda $, where $\lambda
_{s}$ is the wavelength of the acoustic wave and $\lambda $ is the
wavelength for the radiated photon.

Substituting expressions (\ref{FzFx}) into (\ref{Akxdef}) and using the
relation%
\begin{equation}
e^{ib\sin \tau }=\sum_{m=-\infty }^{+\infty }J_{m}\left( b\right) e^{im\tau
},  \label{besseldef}
\end{equation}%
with $J_{m}(b)$ being the Bessel function of the first kind, after the
integration, for the components of the vector potential in the region $z>0$
one finds%
\begin{eqnarray}
A_{x}\left( \omega ,\mathbf{q},z\right) &=&\frac{ee^{i\omega
t_{0}+ik_{z}^{\left( 1\right) }z}}{4\pi ^{2}ck_{z}^{\left( 1\right) }}\tan
\alpha \left[ C_{1}\left( \omega ,\mathbf{q}\right) +2iC_{2}\left( \omega ,%
\mathbf{q}\right) \right] ,  \notag \\
A_{z}\left( \omega ,\mathbf{q},z\right) &=&\frac{ee^{i\omega
t_{0}+ik_{z}^{\left( 1\right) }z}}{4\pi ^{2}ck_{z}^{\left( 1\right) }}\left[
C_{1}\left( \omega ,\mathbf{q}\right) +2i\sqrt{\varepsilon _{1}/\varepsilon
_{0}}C_{2}\left( \omega ,\mathbf{q}\right) \right] .  \label{AxAz}
\end{eqnarray}%
In these expressions we have defined the functions%
\begin{eqnarray}
C_{1}\left( \omega ,\mathbf{q}\right) &=&\frac{e^{ik_{z}^{\left( 0\right)
}l+ia\left[ \sin \left( k_{s}l-\varphi _{1}\right) +\sin \left( \varphi
_{1}\right) \right] }-e^{il\left( \omega /v_{z}-k_{x}\tan \alpha \right) }}{%
\omega /v_{z}-k_{x}\tan \alpha -k_{z}^{\left( 1\right) }},  \notag \\
C_{2}\left( \omega ,\mathbf{q}\right) &=&\sqrt{\frac{k_{z}^{\left( 1\right) }%
}{k_{z}^{\left( 0\right) }}}e^{ia\sin \varphi _{1}}\sum_{m=-\infty
}^{+\infty }J_{m}(a)e^{ilk_{zm}^{\left( 0\right) }/2-im\varphi _{1}}  \notag
\\
&&\times \frac{\sin [(\omega /v_{z}-k_{x}\tan \alpha -k_{zm}^{\left(
0\right) })l/2]}{\omega /v_{z}-k_{x}\tan \alpha -k_{zm}^{\left( 0\right) }},
\label{C12}
\end{eqnarray}%
with $v_{z}=v\cos \alpha $ and%
\begin{equation}
k_{zm}^{\left( 0\right) }=k_{z}^{\left( 0\right) }+mk_{s}.  \label{k3m0}
\end{equation}%
For a special case $\alpha =0$ we have $A_{x}\left( \omega ,\mathbf{q}%
,z\right) =0$ and the expression for $A_{z}\left( \omega ,\mathbf{q}%
,z\right) $ coincides with the formula previously obtained in \cite{Mkrt10}
for the normal incidence.

\section{Radiation intensity}

\label{sec:Radiation}

Given the vector potential we can evaluate the radiation intensity in the
region $z>0$. First we consider the total radiation intensity. The energy
flux through the plane $z=$const is given by the integration of the $z$%
-projection of the Poynting vector:%
\begin{equation}
\frac{c}{4\pi }\int dxdydt[\mathbf{EH}]_{z}=4\pi ^{2}c\int d\omega \int d%
\mathbf{q\,}[\mathbf{E}(\omega ,\mathbf{q},z)\mathbf{H}^{\ast }(\omega ,%
\mathbf{q},z)]_{z},  \label{Sz}
\end{equation}%
where for the Fourier components of the electric and magnetic fields we have
the expressions%
\begin{equation}
\mathbf{H}(\omega ,\mathbf{q},z)=i[\mathbf{k\,A}(\omega ,\mathbf{q},z)],\;%
\mathbf{E}(\omega ,\mathbf{q},z)=-\frac{c}{\omega \varepsilon _{1}}[\mathbf{%
k\,H}(\omega ,\mathbf{q},z)],  \label{HE}
\end{equation}%
with $\mathbf{k}=(k_{x},k_{y},k_{z}^{\left( 1\right) })$. By using these
relations for the spectral-angular density of the radiation intensity we find%
\begin{equation}
I(\omega ,\theta ,\varphi )=\frac{d\mathcal{I}(\omega ,\theta ,\varphi )}{%
d\omega d\theta d\varphi }=4\pi ^{2}\sqrt{\varepsilon _{1}}\frac{\omega ^{2}%
}{c}\sin \theta \cos ^{2}\theta \left\vert \lbrack \mathbf{k\,A}\left(
\omega ,\mathbf{q},z\right) ]\right\vert ^{2},  \label{IntTot}
\end{equation}%
where $\theta $ and $\varphi $ are the polar and azimuthal angles for the
radiated photon wave vector:%
\begin{equation}
\mathbf{k}=\frac{\omega }{c}\sqrt{\varepsilon _{1}}(\sin \theta \cos \varphi
,\sin \theta \sin \varphi ,\cos \theta ).  \label{Photk}
\end{equation}

By taking into account the expressions (\ref{AxAz}) and (\ref{C12}) for the
components of the vector potential, after the averaging over the phase $%
\varphi _{1}$ of particle flight into the plate, for the spectral-angular
density of the radiated energy in the angular region with $\sin \theta <%
\sqrt{\varepsilon _{0}/\varepsilon _{1}}$ we find the expression
\begin{eqnarray}
I\left( \omega ,\theta ,\varphi \right) &=&\frac{e^{2}\sin ^{3}\theta \cos
^{2}\alpha }{\pi ^{2}c\sqrt{\varepsilon _{1}}}\sum_{m=-\infty }^{+\infty
}J_{m}^{2}\left( \frac{\omega \Delta \varepsilon /(2ck_{s})}{\sqrt{%
\varepsilon _{0}-\varepsilon _{1}\sin ^{2}\theta }}\right)  \notag \\
&&\times \left[ \frac{\mathbf{P}(\theta ,\varphi ,\alpha
)}{V(\theta ,\varphi ,\alpha )}-\frac{\mathbf{Q}(\theta ,\varphi
,\alpha )}{U_{m}(\theta ,\varphi ,\alpha )}\frac{\cos ^{1/2}\theta
}{(\varepsilon _{0}/\varepsilon
_{1}-\sin ^{2}\theta )^{1/4}}\right] ^{2}  \notag \\
&&\times \sin ^{2}\left[ \frac{\omega l\sqrt{\varepsilon _{1}}}{2c\cos
\alpha }U_{m}(\theta ,\varphi ,\alpha )\right] .  \label{intentotalgeneral}
\end{eqnarray}%
In (\ref{intentotalgeneral}) we have defined the functions%
\begin{eqnarray}
U_{m}(\theta ,\varphi ,\alpha ) &=&1/\beta _{1}-\sin \theta \cos \varphi
\sin \alpha -\cos \alpha \sqrt{\varepsilon _{0}/\varepsilon _{1}-\sin
^{2}\theta }-\frac{mk_{s}c}{\omega \sqrt{\varepsilon _{1}}}\cos \alpha ,
\notag \\
V(\theta ,\varphi ,\alpha ) &=&1/\beta _{1}-\sin \theta \cos \varphi \sin
\alpha -\cos \theta \cos \alpha ,  \label{UmVdef}
\end{eqnarray}%
and the vectors%
\begin{eqnarray}
\mathbf{P}(\theta ,\varphi ,\alpha ) &=&(\sin \varphi ,\cot \theta \tan
\alpha -\cos \varphi ,-\sin \varphi \tan \alpha ),  \notag \\
\mathbf{Q}(\theta ,\varphi ,\alpha ) &=&(\sqrt{\varepsilon _{1}/\varepsilon
_{0}}\sin \varphi ,\cot \theta \tan \alpha -\sqrt{\varepsilon
_{1}/\varepsilon _{0}}\cos \varphi ,-\sin \varphi \tan \alpha ),
\label{PQdef}
\end{eqnarray}%
with $\beta _{1}=v\sqrt{\varepsilon _{1}}/c$. In general, $\varepsilon _{0}$
and $\varepsilon _{1}$ are functions of $\omega $. In the case of normal
incidence the general formula (\ref{intentotalgeneral}) coincides with the
result previously derived in \cite{Mkrt10}. In the discussion below we will
assume that $\beta _{1}<1$.

Unlike to the case of the normal incidence, in the problem under
consideration we have two different polarizations. For the first one the
electric field for the radiation is parallel to the observation plane (plane
formed by the vectors $\mathbf{k}$ and $\mathbf{v}$) and for the second one
the electric field is perpendicular to the observation plane. These two
polarizations are referred as parallel and perpendicular polarizations,
respectively. The spectral-angular densities for these polarizations we will
denote by $I_{\parallel }\left( \omega ,\theta ,\varphi \right) $ and $%
I_{\perp }\left( \omega ,\theta ,\varphi \right) $. The radiation intensity
for the perpendicular polarization is determined by the $z$-projection of
the corresponding integrated Poynting vector which is given by expression (%
\ref{Sz}) with the replacement $\mathbf{E}(\omega ,\mathbf{q},z)\rightarrow
\mathbf{E}_{\perp }(\omega ,\mathbf{q},z)$, where $\mathbf{E}_{\perp
}(\omega ,\mathbf{q},z)$ is the component of the electric field
perpendicular to the observation plane. Now, by taking into account that
\begin{equation}
\mathbf{\,}[\mathbf{E}_{\perp }(\omega ,\mathbf{q},z)\mathbf{H}^{\ast
}(\omega ,\mathbf{q},z)]_{z}=\frac{\omega ^{2}}{c^{2}}\sqrt{\varepsilon _{1}}%
\cos \theta |\mathbf{A}_{\perp }(\omega ,\mathbf{q},z)|^{2},
\label{PoyntPerp}
\end{equation}%
for the spectral-angular density of the corresponding radiation intensity
one finds%
\begin{equation}
I_{\perp }\left( \omega ,\theta ,\varphi \right) =4\pi ^{2}\frac{\omega ^{4}%
}{c^{3}}\varepsilon _{1}^{3/2}\sin \theta \cos ^{2}\theta |\mathbf{A}_{\perp
}(\omega ,\mathbf{q},z)|^{2}.  \label{Iperp}
\end{equation}%
By making use of expressions (\ref{AxAz}) and (\ref{C12}), after averaging
over the phase $\varphi _{1}$, we get the following final expression%
\begin{eqnarray}
I_{\perp }\left( \omega ,\theta ,\varphi \right) &=&\frac{e^{2}}{4\pi ^{2}c}%
\frac{\left( 1-\sqrt{\varepsilon _{1}/\varepsilon _{0}}\right) ^{2}}{\sqrt{%
\varepsilon _{0}-\varepsilon _{1}\sin ^{2}\theta }}\sin ^{3}\theta \cos
\theta \sin ^{2}\varphi \sin ^{2}(2\alpha )  \notag \\
&&\times \sum_{m=-\infty }^{+\infty }J_{m}^{2}\left( \frac{\omega \Delta
\varepsilon /(2ck_{s})}{\sqrt{\varepsilon _{0}-\varepsilon _{1}\sin
^{2}\theta }}\right) U_{m}^{-2}(\theta ,\varphi )  \notag \\
&&\times \sin ^{2}\left[ \frac{\omega l\sqrt{\varepsilon _{1}}}{2c\cos
\alpha }U_{m}(\theta ,\varphi )\right] .  \label{Iperp1}
\end{eqnarray}%
As we have already noted, the radiation intensity for the perpendicular
polarization vanishes in the case of the normal incidence and this is
explicitly seen from (\ref{Iperp1}). The radiation intensity for the
parallel polarization is found from the relation%
\begin{equation}
I_{\parallel }\left( \omega ,\theta ,\varphi \right) =I\left( \omega ,\theta
,\varphi \right) -I_{\perp }\left( \omega ,\theta ,\varphi \right)
\label{Ipar}
\end{equation}
with the total intensity given by (\ref{intentotalgeneral}).

The radiation intensities for both polarizations have peaks corresponding to
the zeros of the function $U_{m}(\theta ,\varphi ,\alpha )$. The angular
location of the peaks is determined from the equation
\begin{equation}
\sin \theta \cos \varphi \sin \alpha +\cos \alpha \sqrt{\varepsilon
_{0}/\varepsilon _{1}-\sin ^{2}\theta }=1/\beta _{1}-\frac{mk_{s}c}{\omega
\sqrt{\varepsilon _{1}}}\cos \alpha .  \label{Peaks}
\end{equation}%
Note that the location of the peaks does not depend on the amplitude of the
acoustic oscillations (within the framework of the approximation used). The
amplitude will determine the heights of the peaks. The last term in the
right-hand side of (\ref{Peaks}) is induced by the acoustic waves. This term
is of the order $(\omega _{s}/\omega )(c/v_{s})$, with $v_{s}$ being the
velocity for the acoustic waves. In the presence of acoustic waves we have a
set of peaks specified by $m$. The angular location of the peak with $m=0$
coincides with that for the peak in the absence of acoustic waves. The
angular distance between the peaks induced by the acoustic waves is of the
order $(\omega _{s}/\omega )(c/v_{s})$. The condition (\ref{Peaks}) can be
written in a physically more transparent form if, in addition to (\ref{Photk}%
), we introduce the wave vector $\mathbf{k}_{0}$ for the photon inside the
plate (in the absence of the acoustic wave), with the components
\begin{equation}
\mathbf{k}_{0}=\frac{\omega }{c}\sqrt{\varepsilon _{0}}\left( \sin \theta
_{0}\cos \varphi ,\sin \theta _{0}\sin \varphi ,\cos \theta _{0}\right) ,
\label{k0}
\end{equation}%
where $\sin \theta _{0}=\sqrt{\varepsilon _{1}/\varepsilon _{0}}\sin \theta $
(note that the azimuthal angle $\varphi $ is the same inside and outside the
plate). Now it can be seen that, with the help of this vector, the condition
for the peaks is written in the form%
\begin{equation}
\mathbf{k}_{0m}\mathbf{v}=\omega ,\;\mathbf{k}_{0m}\equiv \mathbf{k}_{0}+m%
\mathbf{k}_{s},  \label{peaks2}
\end{equation}%
with $\mathbf{k}_{s}=(0,0,k_{s})$. The latter is the Cherenkov condition for
the medium of the plate. Hence, we conclude that the peaks in the region $%
z>0 $, determined by the condition (\ref{Peaks}), correspond to the
Cherenkov radiation emitted inside the plate and refracted from the
boundary. We can have a situation where the Cherenkov radiation emitted
inside the plate is completely reflected from the boundary and in the
exterior region we have no peaks. In this case the equation (\ref{Peaks})
has no solutions. We can also have cases then the Cherenkov radiation is
confined inside the plate in the absence of the acoustic excitations and the
peaks defined by (\ref{Peaks}) appear as a result of the influence of the
acoustic waves. Note that the location of the peaks can also be controlled
by choosing the incidence angle $\alpha $.

Various special cases of formulas (\ref{intentotalgeneral}) and (\ref{Iperp1}%
) can be considered. In the case of the normal incidence the radiation
intensity for the perpendicular polarization vanishes and for the parallel
polarization we recover the results of Refs. \cite{Grig98} and \cite{Mkrt10}
for the X-ray and optical transition radiations respectively. In the absence
of the acoustic wave we have $\Delta \varepsilon =0$ and in formulas (\ref%
{intentotalgeneral}) and (\ref{Iperp1}) the $m=0$ term contributes only. In
this case we obtain the quasi-classical approximation for the radiation
intensity for oblique incidence. The corresponding exact expression for the
radiation intensity in this problem is well known from the literature \cite%
{Pafo69} (see also Refs. \cite{TerMik}-\cite{Poty09}). The features of the
optical transition radiation in a finite thickness plate for an oblique
incidence have been discussed in Refs. \cite{Ruzi86} on the base of
Pafomov's formulas. For a transparent material in the over-threshold case
and under the condition $l\omega /v\gg 1$, the dominant contribution comes
from the term in the exact formula with the resonant factor $\sin
^{2}(y)/y^{2}$, with $\ y$ given by the argument of the sin function in (\ref%
{intentotalgeneral}) with $m=0$. Now, for simplicity considering the case of
the radiation into the vacuum, it can be seen that for a relativistic
particle with $1-v/c\ll 1$, the radiation intensity near the Cherenkov peaks
is well approximated by the formulas obtained from (\ref{intentotalgeneral})
and (\ref{Iperp1}) in the limit $\Delta \varepsilon =0$.

For given values of $\alpha $ and $\varphi $, the equation (\ref{Peaks})
determines the location of the peaks with respect to the polar angle $\theta
$. We will denote the corresponding values by $\theta ^{(m)}$. At the peaks
the total radiation intensity is given by the expressions%
\begin{eqnarray}
&& I(\omega ,\theta ^{(m)},\varphi ) =\frac{e^{2}\varepsilon
_{1}l^{2}\omega ^{2}\sin ^{3}\theta ^{(m)}\cos \theta ^{(m)}}{4\pi
^{2}c^{3}(\varepsilon
_{0}-\varepsilon _{1}\sin ^{2}\theta ^{(m)})^{1/2}}  \notag \\
&& \qquad \times \mathbf{Q}^{2}(\theta ^{(m)},\varphi ,\alpha )J_{m}^{2}\left( \frac{%
\omega \Delta \varepsilon /(2ck_{s})}{\sqrt{\varepsilon _{0}-\varepsilon
_{1}\sin ^{2}\theta ^{(m)}}}\right) .  \label{Ipeak1}
\end{eqnarray}%
For the relative contribution of the component with the perpendicular
polarization one gets
\begin{equation}
\frac{I_{\perp }\left( \omega ,\theta ^{(m)},\varphi \right) }{I\left(
\omega ,\theta ^{(m)},\varphi \right) }=\left( 1-\sqrt{\varepsilon
_{1}/\varepsilon _{0}}\right) ^{2}\frac{\sin ^{2}\varphi \sin ^{2}\alpha }{%
\mathbf{Q}^{2}(\theta ^{(m)},\varphi ,\alpha )}.  \label{IpIpeak}
\end{equation}%
In the absence of the ultrasonic vibrations the location of the peak is
given by $\theta ^{(0)}$. The ultrasound reduces the height of this peak by
the factor $J_{0}^{2}\left( \omega \Delta \varepsilon /(2ck_{s}\sqrt{%
\varepsilon _{0}-\varepsilon _{1}\sin ^{2}\theta ^{(0)}})\right) $. In
particular, for a given radiation frequency, the frequency or the amplitude
of the ultrasound can be tuned to eliminate this peak (see the numerical
example below).

In the limit $l\rightarrow \infty $, by using the formula $%
\lim_{x\rightarrow \infty }\sin ^{2}(ux)/x=\pi u^{2}\delta (u)$, with $%
\delta (u)$ being the Dirac delta function, for the spectral-angular density
of the total radiation energy per unit length we find%
\begin{eqnarray}
&& I_{\infty }\left( \omega ,\theta ,\varphi \right)
=\frac{e^{2}\omega \sin ^{3}\theta \cos \theta \cos \alpha }{2\pi
c^{2}\sqrt{\varepsilon _{0}/\varepsilon _{1}-\sin ^{2}\theta
}}\delta \left[ U_{m}(\theta ,\varphi
,\alpha )\right]  \notag \\
&& \qquad \times \mathbf{Q}^{2}(\theta ,\varphi ,\alpha
)\sum_{m=-\infty }^{+\infty
}J_{m}^{2}\left( \frac{\omega \Delta \varepsilon /(2ck_{s})}{\sqrt{%
\varepsilon _{0}-\varepsilon _{1}\sin ^{2}\theta }}\right) ,  \label{Iominf}
\end{eqnarray}%
where $I_{\infty }\left( \omega ,\theta ,\varphi \right) =\lim_{l\rightarrow
\infty }I\left( \omega ,\theta ,\varphi \right) /l$. A similar formula is
obtained from (\ref{Iperp1}) for the perpendicular polarization. In this
case the radiation along a given direction has discrete spectrum $\omega
=\omega _{m}$, determined by the relation $U_{m}(\theta ,\varphi ,\alpha )=0$%
, or explicitly%
\begin{equation}
\omega _{m}=\frac{|m|k_{s}v\cos \alpha }{|1-\beta _{1}\sin \theta \cos
\varphi \sin \alpha -\beta _{1}\cos \alpha \sqrt{\varepsilon
_{0}/\varepsilon _{1}-\sin ^{2}\theta }|}.  \label{omegam}
\end{equation}%
The spectral distribution of the radiation intensity is obtained from (\ref%
{Iominf}) by the integration over $\theta $ and $\varphi $.

\section{Radiation in a plate of fused quartz}

\label{sec:Numeric}

Having the general analysis for the radiation intensity, we turn to
numerical examples. We will be interested in the optical transition
radiation. For the numerical evaluation of the radiation intensity the
material of the plate should be specified. We assume that the plate is made
of fused quartz with the velocity of longitudinal ultrasonic vibrations $%
v_{s}\approx 5.6\times 10^{5}\;$cm/s. For the corresponding dielectric
permittivity we use the Sellmeier dispersion formula%
\begin{equation}
\varepsilon _{0}=1+\sum_{i=1}^{3}\frac{a_{i}\lambda ^{2}}{\lambda
^{2}-l_{i}^{2}}  \label{sellmeier}
\end{equation}%
with the parameters $a_{1}=0.6961663$, $a_{2}=0.4079426$, $a_{3}=0.8974794$,
$l_{1}=0.0684043$, $l_{2}=0.1162414$, $l_{3}=9.896161$. In (\ref{sellmeier}%
), $\lambda $ is the wavelength of the radiation measured in micrometers.
Formula (\ref{sellmeier}) well describes the dispersion properties of fused
quartz in the range $0.2\mathrm{\mu m}\leqslant \lambda \leqslant 6.7\mathrm{%
\mu m}$. In this spectral range fused quartz is very weakly absorbing.

In figures below we plot the spectral-angular density of the total
radiation intensity in the forward direction, $I(\omega ,\theta
,\varphi )/\hbar $, and the
spectral-angular density for the component with perpendicular polarization, $%
I_{\perp }(\omega ,\theta ,\varphi )/\hbar $, for electrons with
the energy $2$ MeV and for the plate thickness $l=1$ cm. For the
oscillation amplitude we have taken the value $\Delta
n/n_{0}=0.05$, where $n_{0}$ is the number of electrons per unit
volume for fused quartz.

As it is seen from the graphs and in accordance with general
features described above, the presence of the acoustic wave leads
to the appearance of new peaks in both angular and spectral
distributions of the radiation intensity. The height of the peaks
can be tuned by choosing the parameters of the acoustic wave. In
particular, the peak in the radiation intensity which is present
in the absence of the acoustic wave is reduced by the factor
$J_{0}^{2}(a)$, where $a$ is the argument of the Bessel function
in (\ref{intentotalgeneral}). In particular, this peak can be
completely
removed by taking the parameters of the acoustic wave in such a way to have $%
a=j_{0,s}$, $s=1,2,3,\ldots $, where $z=j_{0,s}$ are the zeroes of the
function $J_{0}(z)$.

\begin{figure}[tbph]
\begin{center}
\begin{tabular}{cc}
\epsfig{figure=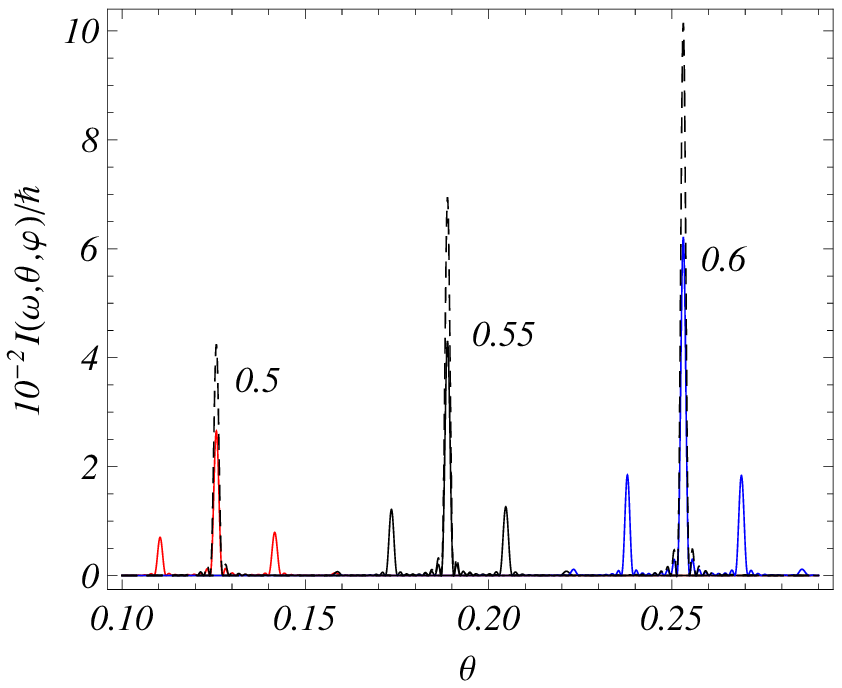,width=7.cm,height=6.cm} & \quad %
\epsfig{figure=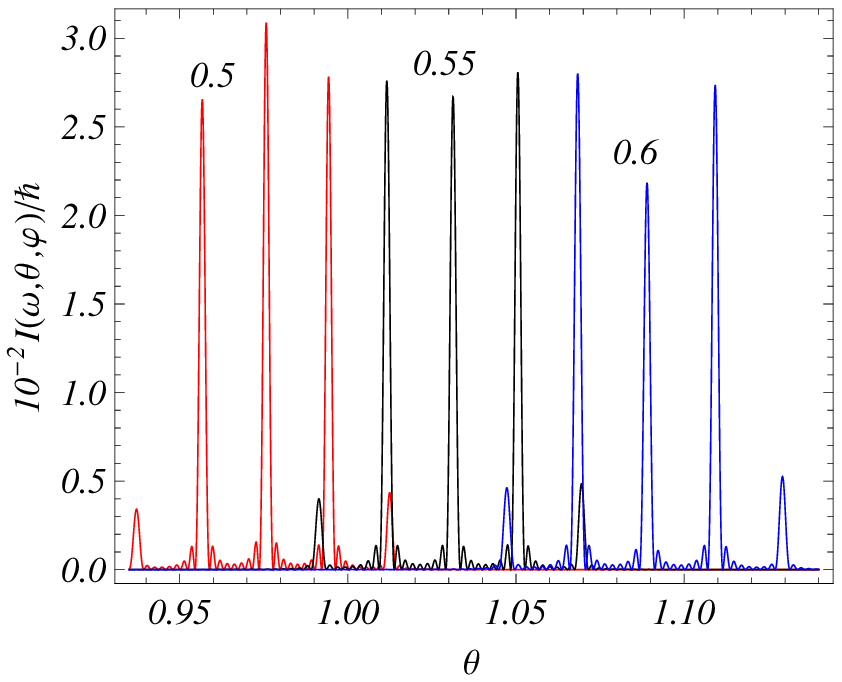,width=7.cm,height=6.cm}%
\end{tabular}%
\end{center}
\caption{The angular distributions of the radiation intensity for two
separate sets of peaks for different values of the incidence angle $\protect%
\alpha $ (numbers near the curves). The dashed curves correspond to the
transition radiation in the case where the acoustic wave is absent. The
values of the parameters are as follows: $\protect\omega =2.73\times 10^{14}
$ Hz, $\protect\varphi =0.369$, $\protect\nu _{s}=5$ MHz. }
\label{fig1}
\end{figure}

\begin{figure}[tbph]
\begin{center}
\begin{tabular}{cc}
\epsfig{figure=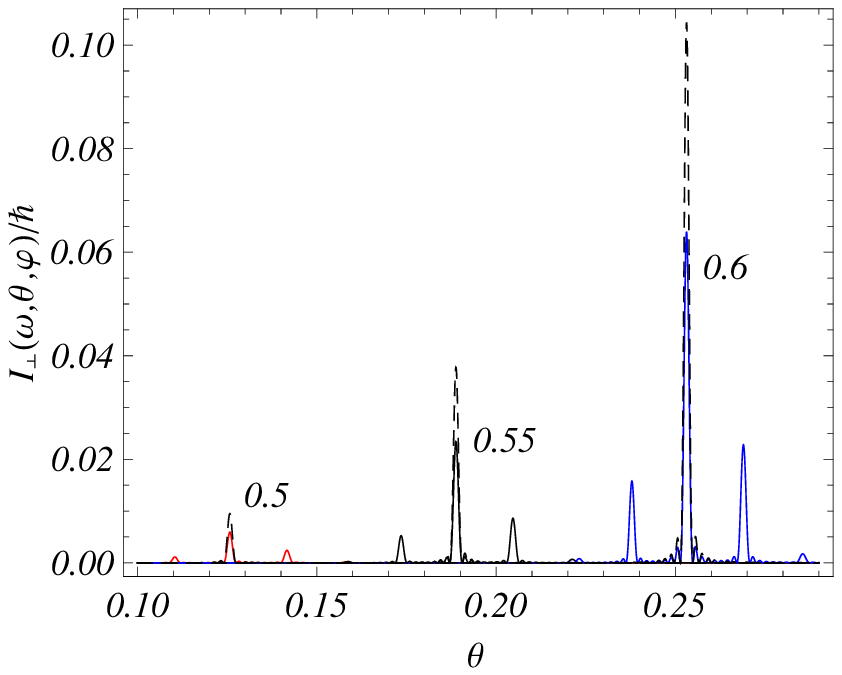,width=7.cm,height=6.cm} & \quad %
\epsfig{figure=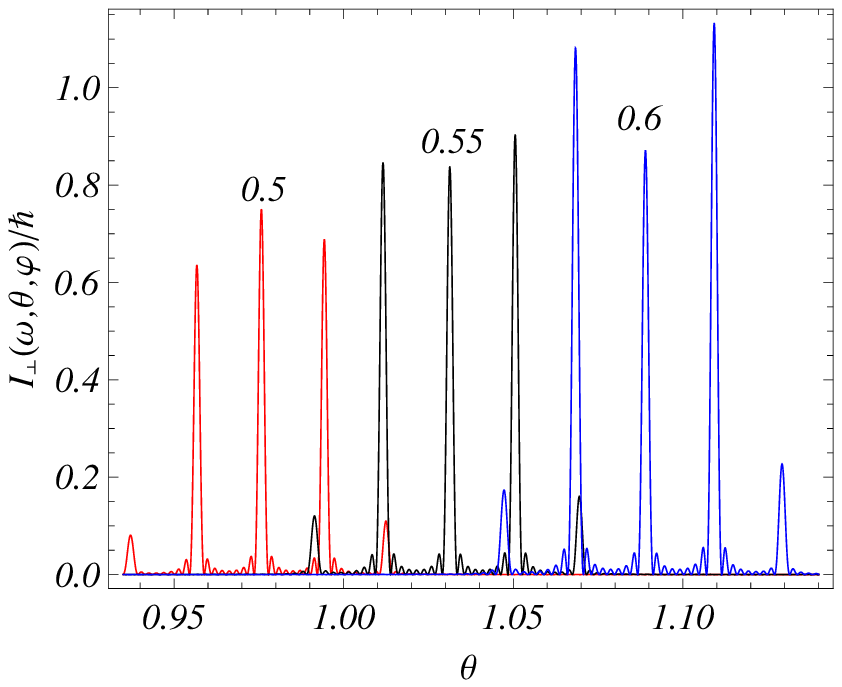,width=7.cm,height=6.cm}%
\end{tabular}%
\end{center}
\caption{The same as in figure \protect\ref{fig1} for the perpendicular
polarization. }
\label{fig2}
\end{figure}

\begin{figure}[tbph]
\begin{center}
\begin{tabular}{cc}
\epsfig{figure=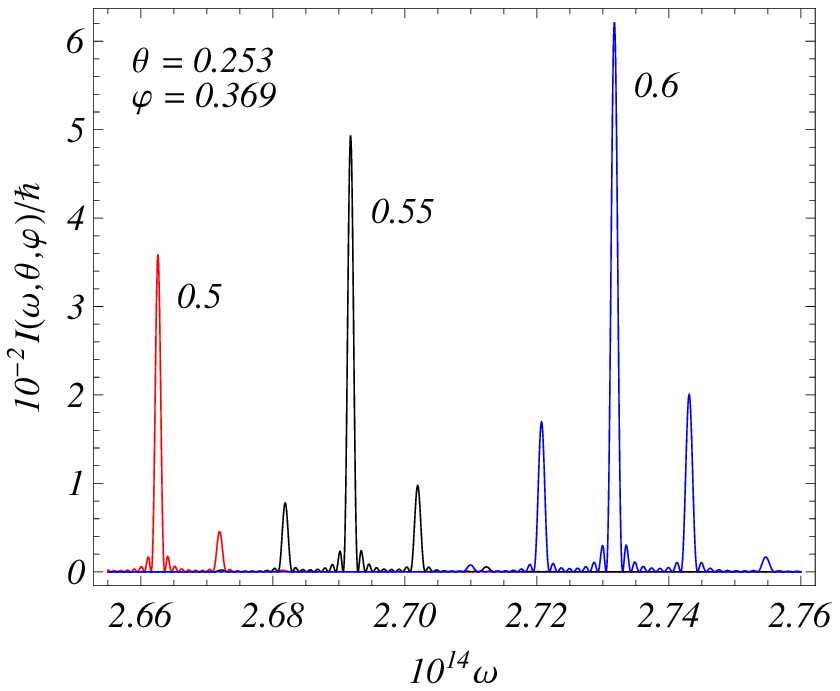,width=7.cm,height=6.cm} & \quad %
\epsfig{figure=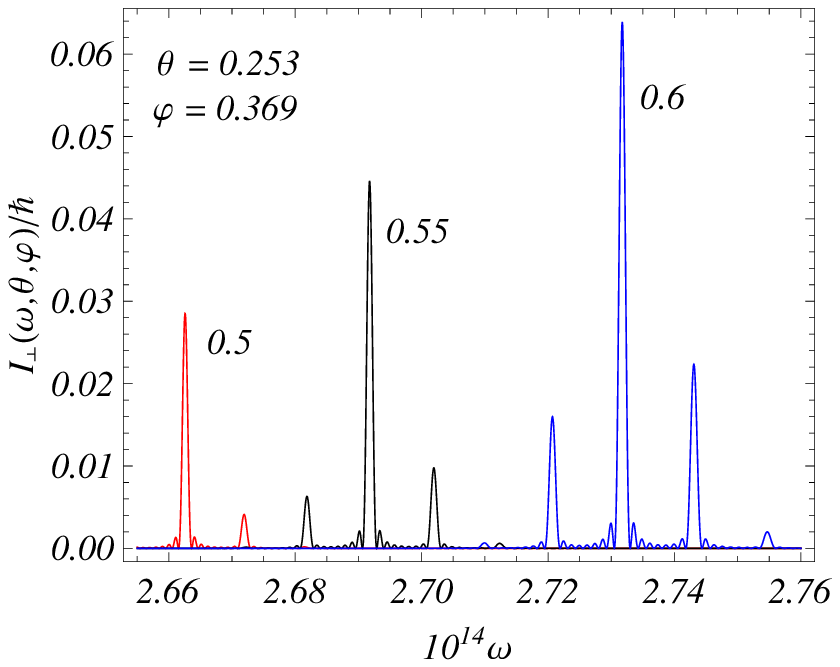,width=7.cm,height=6.cm}%
\end{tabular}%
\end{center}
\caption{The spectral distribution of the radiation intensity for a polar
angle $\protect\theta =0.253$ for the total intensity (left panel) and for
the radiation with perpendicular polarization (right panel). The values of
the other parameters are the same as those for figure \protect\ref{fig1}.}
\label{fig3}
\end{figure}

\begin{figure}[tbph]
\begin{center}
\begin{tabular}{cc}
\epsfig{figure=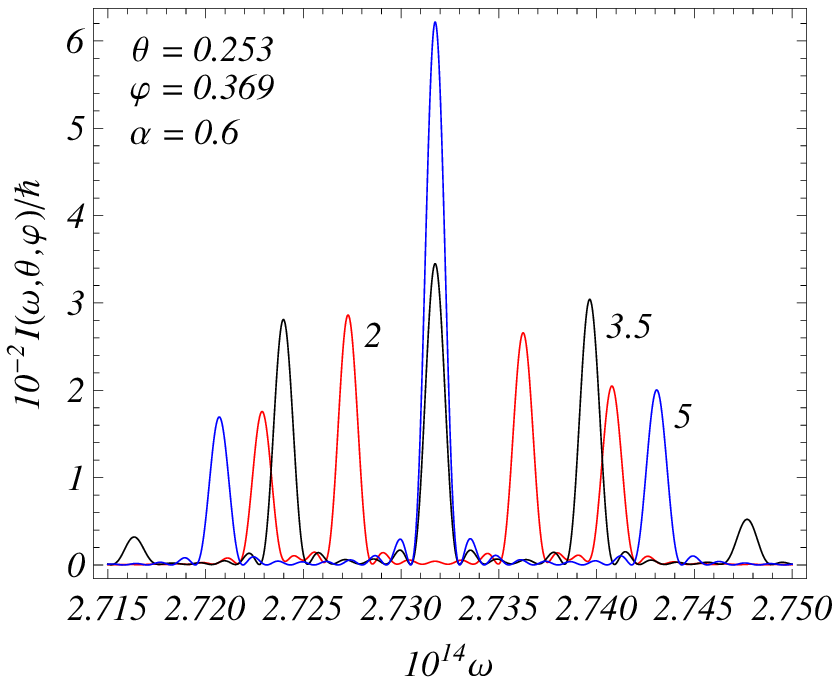,width=7.cm,height=6.cm} & \quad %
\epsfig{figure=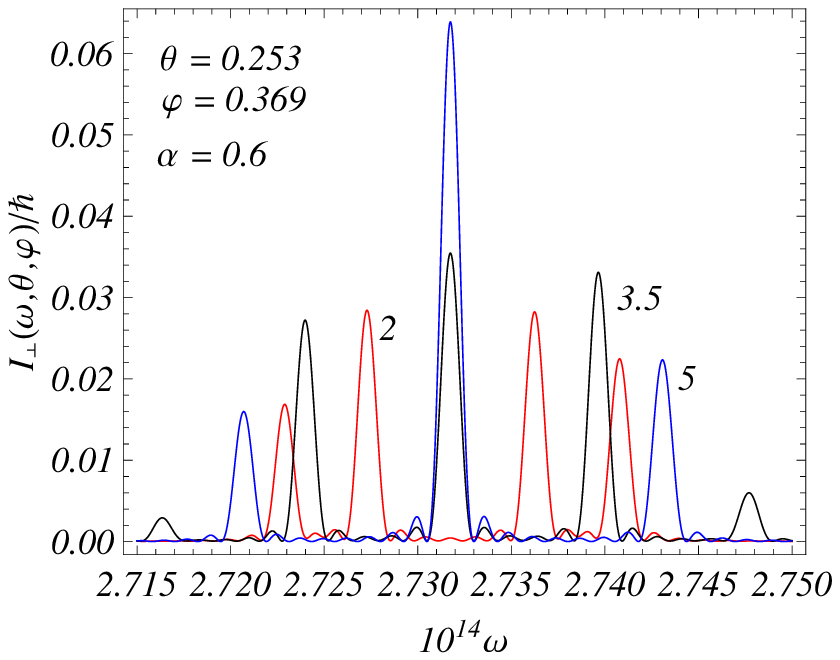,width=7.cm,height=6.cm}%
\end{tabular}%
\end{center}
\caption{Spectral distribution for the total intensity (left panel) and for
the intensity of the radiation with perpendicular polarization (right panel)
for separate values of the acoustic wave frequency $\protect\nu _{s}$
(numbers near the curves).}
\label{fig4}
\end{figure}

In figure \ref{fig1} we plot the radiation intensity, defined by (\ref%
{intentotalgeneral}), as a function of the polar angle $\theta $ for
separate values of the angle $\alpha $ (numbers near the curves) and for the
cyclic frequency $\omega =2.73\times 10^{14}$ Hz. For the azimuthal angle we
have taken the value $\varphi =0.369$. The graphs in figure \ref{fig1} are
plotted for the frequency of acoustic wave $\nu _{s}=5$ MHz. Dashed curves
correspond to the radiation in the situation where the acoustic wave is
absent. Left and right panels show two different sets of the peaks. Similar
graphs for the radiation with the perpendicular polarization are displayed
in figure \ref{fig2} for the same values of the parameters. The peaks in the
spectral distribution of the radiation intensity for a polar angle $\theta
=0.253$ are shown in figure \ref{fig3} for the total intensity (left panel)
and for the radiation with perpendicular polarization (right panel). The
values of the other parameters are the same as those for figure \ref{fig1}.
In figure \ref{fig4} we show the spectral distribution for the total
intensity (left panel) and for the intensity of the radiation with
perpendicular polarization (right panel) for separate values of the acoustic
wave frequency $\nu _{s}$ (numbers near the curves). An example is presented
where the $m=0$ peak in the absence of the acoustic wave is completely
suppressed by the acoustic wave.

\section{Conclusion}

\label{sec:Conc}

We have investigated the transition radiation under oblique incidence of a
charged particle in the presence of acoustic waves. In the quasi-classical
approximation, formulas are derived for vector potential of the
electromagnetic field and for the radiation intensity in the forward
direction. The spectral-angular density of the radiated energy is given by
formula (\ref{intentotalgeneral}) for the total radiation and by (\ref%
{Iperp1}) for the component with perpendicular polarization. The radiation
intensities for both polarizations have strong peaks with the angular
location determined from the condition (\ref{Peaks}). This condition is
written in a physically more transparent form, Eq. (\ref{peaks2}), in terms
of the photon wave vector inside the plate. The latter clearly shows that
the peaks correspond to the Cherenkov radiation emitted inside the plate and
refracted from the boundary. In the presence of acoustic waves we have a set
of peaks specified by $m$. The angular distance between them is of the order
$(\omega _{s}/\omega )(c/v_{s})$. The radiation intensity at the peaks is
given by expressions (\ref{Ipeak1}) and (\ref{IpIpeak}). The
angular-frequency parameters of the peaks can be controlled by tuning the
amplitude and the wavelength of the ultrasound. In particular, we can have a
situation where the peaks in the forward direction are absent in the absence
of the acoustic excitations and they appear as a result of the influence of
the acoustic waves. The numerical examples are given for the optical
transition radiation in a plate of fused quartz. These results show that the
acoustic waves allow to control the both angular and spectral parameters of
the radiation. In particular, new resonance peaks appear in the
spectral-angular distribution of the radiation intensity. An additional
parameter which can be used for the control is the incidence angle.

\section{Acknowledgment}

The authors are grateful to Professors Babken Khachatryan and
Levon Grigoryan for valuable discussions and suggestions.


\begin{thebibliography}{99}
\bibitem{TerMik} M. L. Ter-Mikaelian, \textit{High Energy Electromagnetic
Processes in Condensed Media} (Wiley Interscience, New York, 1972).

\bibitem{Gari83} G.M. Gharibian, S. Yan, \textit{Rentgenovskoye Perekhodnoye
Izluchenie} (Izd. AN Arm. SSR, Yerevan, 1983, in Russian).

\bibitem{Ginz84} V. L. Ginzburg and V. N. Tsytovich, \textit{Transition
Radiation and Transition Scattering} (Adam Hilger, Bristol, 1990).

\bibitem{Baie89} V. N. Baier, V. M. Katkov, V. M. Strakhovenko, \textit{%
Electromagnetic Processes at High Energies in Oriented Single Crystals}
(World Scientific, Singapore, 1998).

\bibitem{Rull98} P. Rullhusen, X. Artru, P. Dhez, \textit{Novel Radiation}
\textit{Sources Using Relativistic Electrons} (World Scientific, Singapore,
1998).

\bibitem{Poty09} A. P. Potylitsin, \textit{Izluchenie Electronov v
Periodicheskikh Structurakh} (Izdatelstvo Nauchno-Tekhnicheskoi Literaturi,
Tomsk, 2009, in Russian).

\bibitem{MkrtDR} A. R. Mkrtchyan, L. Sh. Grigoryan, A. A. Saharian, A. N.
Didenko, Acustica \textbf{75}, 1984 (1991); A. A. Saharian, A. R. Mkrtchyan,
L. A. Gevorgian, L. Sh. Grigoryan, B. V. Khachatryan, Nucl. Instr. Meth. B
\textbf{173}, 211 (2001).

\bibitem{Mkrt91} A. R. Mkrtchyan, H. A. Aslanyan, A. H. Mkrtchyan, R. A.
Gasparyan, Phys. Lett. A \textbf{152}, 297 (1991).

\bibitem{Mkrt86} A. R. Mkrtchyan, R. A. Gasparyan, R. G. Gabrielyan, Phys.
Lett. A \textbf{115}, 410 (1986); JETP \textbf{93} 432 (1987);
Phys. Lett. A \textbf{126}, 528 (1988); L. Sh. Grigoryan et al.,
Nucl. Instr. Meth. B \textbf{173}, 132 (2001); \textbf{173}, 184
(2001); L. Sh. Grigoryan, A. H. Mkrtchyan, H. F. Khachatryan, V.
U. Tonoyan, W. Wagner, Nucl. Instr. Meth. B \textbf{201}, 25
(2003); L. Sh. Grigoryan et al., Nucl. Instr. Meth. B
\textbf{212}, 51 (2003).

\bibitem{Mkrt02} A. R. Mkrtchyan, A. A. Saharian, L. Sh. Grigoryan, B. V.
Khachatryan, Mod. Phys. Lett. A \textbf{17}, 2571 (2002); A. R. Mkrtchyan,
A. A. Saharian, V. V. Parazian, Mod. Phys. Lett. B \textbf{20}, 1617 (2006).

\bibitem{Saha04Brem} A. A. Saharian, A. R. Mkrtchyan, V. V. Parazian, L. Sh.
Grigoryan, Mod. Phys. Lett. A \textbf{19}, 99 (2004); A. R. Mkrtchyan, A. A.
Saharian, V. V. Parazian, Mod. Phys. Lett. B \textbf{23}, 2573 (2009).

\bibitem{Koro04} A. V. Korol, A. V. Solov'yov and W. Greiner, Int. J. Mod.
Phys. E \textbf{13}, 867 (2004).

\bibitem{Grig98} L. Sh. Grigoryan, A. H. Mkrtchyan, A. A. Saharian, Nucl.
Instr. Meth. B \textbf{145}, 197 (1998).

\bibitem{Mkrt10} A. R. Mkrtchyan, V. V. Parazian, A. A. Saharian, Mod. Phys.
Lett. B \textbf{24}, 2693 (2010).

\bibitem{Grig09} L. Sh. Grigoryan, A. R. Mkrtchyan, H. F. Khachatryan, S. R.
Arzumanyan, W. Wagner, J. Phys.: Conference Series \textbf{236}, 012012
(2010); arXiv:0911.3494.

\bibitem{Khachatryan} B. V. Khachatryan, Doctor of Sciences Thesis. Yerevan,
2004 (in Russian).

\bibitem{Pafo69} V. E. Pafomov, Proc. FIAN SSSR, Nuclear Physics and Particle
Interaction with Matter, Vol. 44, Nauka, Moscow, 1969, p. 90.

\bibitem{Ruzi86} J. Ru\v{z}i\v{c}ka, J. Mehes, Nucl. Instr. Meth. A \textbf{%
250}, 491 (1986); A. Hrmo, J. Ruzicka, Nucl. Instrum. Methods A \textbf{451}%
, 506 (2000).
\end{thebibliography}
\end{document}